# Impact of indirect transitions on valley polarization in $WS_2$ and $WSe_2$

Rasmus H. Godiksen[1], Shaojun Wang[1,2], T.V. Raziman[1],

Jaime Gómez Rivas[1], Alberto G. Curto[1,3,4*]

*[1]Dep. Applied Physics and Institute for Photonic Integration,*

*Eindhoven University of Technology, Eindhoven, The Netherlands*

*[2]MOE Key Lab. of Modern Optical Technologies and Jiangsu Key Lab. of Advanced Optical Manufacturing Technologies, School of Optoelectronic Science and Engineering, Soochow University, Suzhou 215006, China*

*[3]Photonics Research Group, Ghent University-imec, Ghent, Belgium*

*[4]Center for Nano- and Biophotonics, Ghent University, Ghent, Belgium*

* Corresponding author: A.G.Curto@TUe.nl

**Controlling the momentum of carriers in semiconductors, known as valley polarization, is a new resource for optoelectronics and information technologies. Materials exhibiting high polarization are needed for valley-based devices. Few-layer $WS_2$ shows a remarkable spin-valley polarization above 90%, even at room temperature. In stark contrast, polarization is absent for few-layer $WSe_2$ despite the expected material similarities. Here, we explain the origin of valley polarization in both materials based on the interplay between two indirect optical transitions. We show that the relative energy minima at the Λ- and K-valleys in the conduction band determine the spin-valley polarization of the direct K-K transition. Polarization appears as the energy of the K-valley rises above the Λ-valley as a function of temperature and number of layers. Our results advance the understanding of the high spin-**

**valley polarization in $WS_2$. This insight will impact the design of both passive and tunable *valleytronic* devices operating at room temperature.**

Transition metal dichalcogenides (TMDs) such as $MoS_2$, $WS_2$, or $WSe_2$ are layered semiconductors with unique spin-valley physics. The coupling between spin and momentum for excited carriers opens a new path to access the valley degree of freedom.[1] Valley polarization arises from a difference in exciton populations at the K- and K'-points of the hexagonal Brillouin zone (Figure 1a)[1–3], where local energy minima known as valleys lie. At these diametrically opposite points in reciprocal space, strong spin-orbit splitting occurs in the top valence bands. The different signs of the splitting in the K- and K'-valleys[4] result in the coupling of spin and valley indexes and leads to spin-dependent optical and electronic properties. The K- and K'-valleys can be selectively excited using right- or left-handed circularly polarized light.[5] TMDs constitute thus a fascinating platform for future *valleytronic*[6], optoelectronic[7–9], and nanophotonic[10,11] devices exploiting the spin, valley, and layer indexes.

The layered nature of TMDs enables a high degree of control over valley polarization. A monolayer possesses a direct band gap, whereas in the few-layer regime the band gap becomes indirect.[12–14] Light emission in few-layer TMDs is dominated by indirect transitions from the Λ- and K-points to the Γ-point in the band structure (Figure 1a). These indirect transitions are typically unpolarized. At higher energy, polarized intravalley transitions with direct character can still occur in the K- or K'-valleys.[15] The degree of circular polarization can be used as a proxy for valley polarization. It is defined as $\mathrm{DOCP} = (I_{\sigma_+} - I_{\sigma_-})/(I_{\sigma_+} + I_{\sigma_-})$, where $I_{\sigma_+}$ and $I_{\sigma_-}$ are the photoluminescence intensities with right- and left-handed circular polarization, respectively.

Valley polarization can reach values near unity at cryogenic temperatures for monolayer $MoS_2$.[1] With increasing temperature, however, the initial polarization quickly depolarizes due to intervalley scattering between the K- and K'-points[16], limiting applications at room temperature. At higher temperatures, a valley polarization enhancement for a monolayer has been realized through interaction with graphene[17,18], reaching up to 40% DOCP for graphene-encapsulated $WS_2$.[19] In contrast to the monolayer case, the DOCP reaches 65% for bilayer $WS_2$ even at room temperature.[20] Such a large spin-valley polarization in bilayer $WS_2$ is not well understood yet.[21,22]

Despite sharing several properties with $WS_2$ due to the common W atom, valley polarization is absent in $WSe_2$ at room temperature. This discrepancy between bilayer $WS_2$ and $WSe_2$, illustrated in Figures 1c and 1d, is inconstent with theoretical predictions.[15] Hence, it remains unresolved, as stated by Bussolotti *et al*.[21] An understanding of the spin-valley properties that lead to high and low polarization in $WS_2$ and $WSe_2$ is, therefore, essential for practical applications at room temperature.[23] Gaining insight into the spin-valley physics of bilayer TMDs would also be beneficial for spin-layer locking effects[24–27], layer-dependent spin relaxation[28], and the spin-valley Hall effect in few-layer systems.[29,30]

Here, we demonstrate the critical role of the Λ-valley on the spin-valley polarization in few-layer $WS_2$ and $WSe_2$ through a combined investigation of polarization- and temperature-resolved photoluminescence (PL). By varying both the number of layers and the temperature, we analyze the interplay between the momentum-allowed direct transition (K-K) and two momentum-forbidden indirect transitions (K-Γ and Λ-Γ). We find that a change in the dominant indirect transition channel with temperature determines the observation of spin-valley polarization. In bilayer $WSe_2$, we reveal the existence of a crossover temperature at which the dominant indirect transition switches from K-Γ to Λ-Γ as the Λ-point energy shifts lower in energy than the K-point. Below this crossover temperature, the polarization of the direct K-K transition begins to increase even for highly off-resonant excitation. We demonstrate the dependence of the valley polarization of the direct K-K transition on the K-Λ energy difference in the conduction band. In contrast to $WSe_2$, the Λ-Γ indirect transition dominates the emission in $WS_2$ resulting in high polarization even at room temperature. Based on our results, we explain how both temperature and number of layers affect spin-valley polarization in $WS_2$ and $WSe_2$. Therefore, we identify a missing piece of the puzzle for understanding and achieving high spin-valley polarization in few-layer semiconductors.

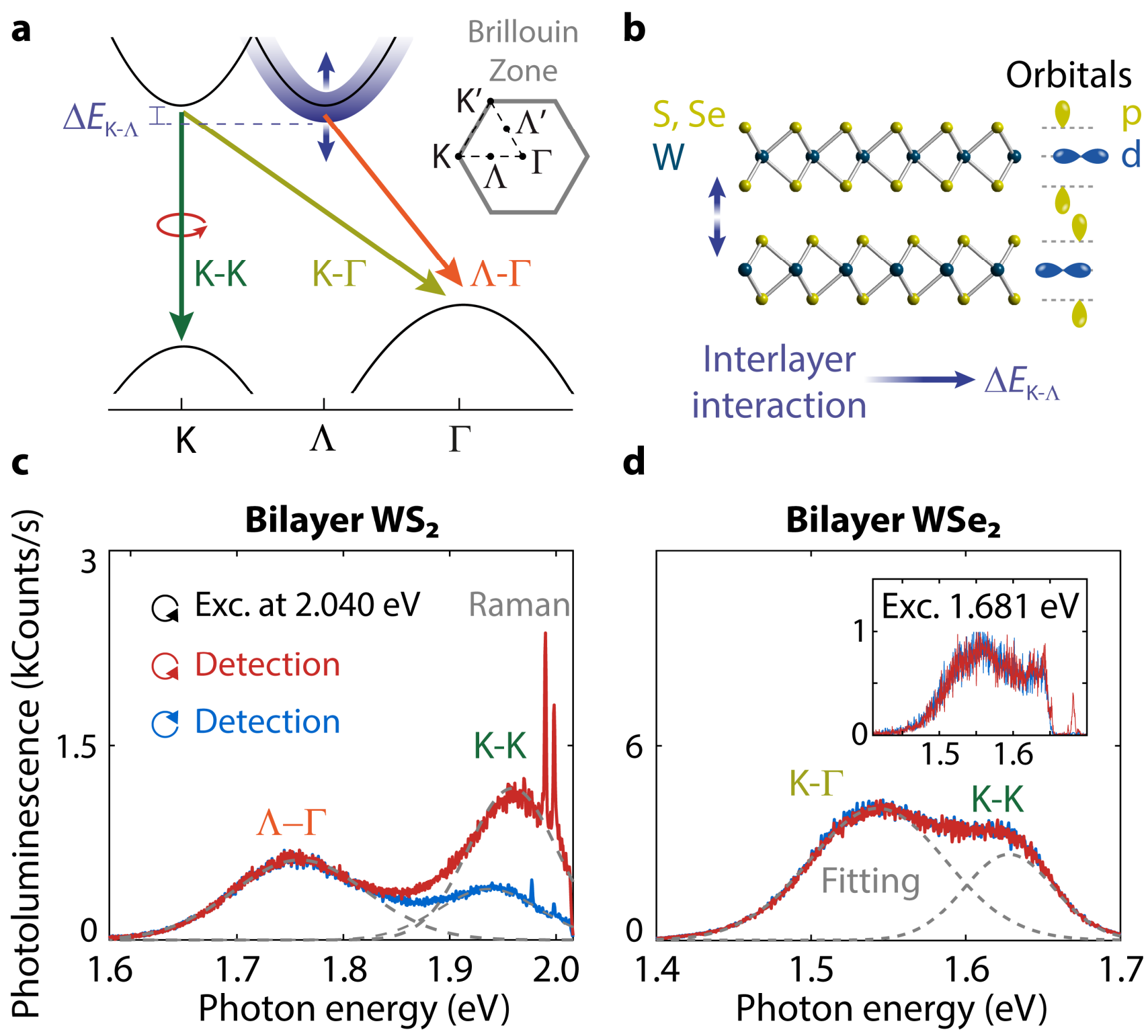

**Figure 1 | Direct and indirect optical transitions in few-layer TMDs and their spin-valley polarization. a,** Possible exciton transitions in few-layer $WS_2$ and $WSe_2$. Two indirect transitions can occur depending on whether the conduction band minimum is either at the K- or the Λ-valley.

Their relative energy difference is $\Delta E_{K\text{-}\Lambda}$. Intervalley K-K' scattering is omitted for clarity. **b,** Illustration of a bilayer $WS_2$ or $WSe_2$ and the orbital characters of the K- and Λ-valleys. The interlayer interactions through the *p*-orbitals control the Λ-point energy resulting in changes of $\Delta E_{K\text{-}\Lambda}$. **c-d,** Polarization-resolved photoluminescence spectra for bilayer $WS_2$, and bilayer $WSe_2$ excited with photon energy 2.040 eV at room temperature. There is high polarization in $WS_2$ and no polarization in $WSe_2$. Inset: lack of polarization for bilayer $WSe_2$ even when excited with 1.681 eV close to resonance to its K-K transition.

**Spin-valley polarization in few-layer $WS_2$ and $WSe_2$**

First, we consider the typical band structure of a bilayer TMD (Figure 1a). The bilayer band gap is indirect because the valence band maximum shifts from K to Γ from one to two layers. The nature of the indirect transition depends on the competition between the Λ and K conduction band energy minima (orange and light green arrows in Figure 1a). To study the impact of these indirect transitions on polarization, we will exploit their dependence on temperature and number of layers to tune the band structure.

The effect of interlayer interactions on the band structure is highly dependent on momentum, leading to a different layer and temperature dependence for the energy of the K-K, K-Γ, and Λ-Γ transitions (Table 1).[16,31] At the K-point, *d*-orbitals from the transition metals determine the top-most band structure.[5] Increasing the temperature expands the covalent bond length between the atoms reducing the energy gap at the K-point. The transition metal atoms are protected between the chalcogens, which results in insensitivity of the K-point to the surrounding medium and, therefore, to the number of layers. On the other hand, the chalcogen atoms lie close to both the surrounding medium and the adjacent layers. The chalcogen *p*-orbitals that dominate at the Λ-point extend outside the atomic plane, rendering it sensitive to interlayer interactions (Figures 1b) .[31] With increasing temperature, the out-of-plane *p*-orbitals extend in length and come closer to each other, thereby increasing their interaction because the interlayer distance due to van der Waals forces between the layers is not temperature dependent. Consequently, the Λ-valley increases in energy with increasing temperature. Increasing the number of layers, on the contrary, results in a decrease of the Λ-valley energy because more out-of-plane *p*-orbitals interact with neighboring layers. As summarized in Table 1, we can utilize both temperature and the number of layers to alter the direct and indirect transitions of $WS_2$ and $WSe_2$.

**Table 1 |** The number of layers and temperature affect the band structure of few-layer $WS_2$ and $WSe_2$, resulting in different dependences for the transition energies between different points in momentum space.

| Energy | Increase in #*L* | Increase in *T* |
|---|---|---|
| $E_{K\text{-}K}$ | Near constant | Decreases |
| $E_{\Lambda\text{-}\Gamma}$ | Decreases | Increases |
| $E_{K\text{-}\Gamma}$ | Decreases | Decreases |

To compare the valley polarization of $WS_2$ and $WSe_2$, we excite our samples with circularly polarized light and measure the polarization of the emission with a circular polarization analyzer and a spectrometer (see Methods). We observe a stark difference in circular polarization for bilayer $WS_2$ and $WSe_2$ (compare high and low values in Figures 1c and 1d) for excitation with laser photon energy of 2.040 eV, which is close to resonance for $WS_2$. We confirmed that the low polarization for $WSe_2$ is not due to off-resonant excitation by using two additional excitation energies of 1.796 eV and 1.681 eV (Figure 1d, inset and Supplementary Figure 1). We still observed no polarization at room temperature despite having nearly the same detuning with the K-K emission of 66 meV in $WSe_2$ excited by 1.681 eV compared to $WS_2$ excited by 2.040 eV. In this work, we will demonstrate the dependence of the polarization in $WS_2$ and $WSe_2$ on the indirect band gap character controlled by the energy difference $\Delta E_{K\text{-}\Lambda}$ (Figure 1a). To clarify the role played by $\Delta E_{K\text{-}}$ $_{\Lambda}$ on the differences and similarities between the polarization of $WS_2$ and $WSe_2$, we measure next the changes in spectra and polarization as a function of the number of layers and temperature.

**The role of the indirect optical transitions in polarization**

We prepare samples with varying numbers of layers of $WS_2$ and $WSe_2$. We measure their PL spectra and determine the position of the direct and indirect transition peaks (Supplementary Figure 2). The peak energies as a function of the number of layers show that the separation between the direct and indirect peaks increases faster with thickness for $WS_2$ compared to $WSe_2$ (Figures 2a and 2b). This difference is a consequence of the origin of their indirect emission, which stems from Λ-Γ transitions for $WS_2$ while it originates from K-Γ transitions for $WSe_2$ at room temperature.[31] The larger increase in energy shift with thickness for the Λ-Γ transition is due to the larger impact of interlayer interactions on the Λ-valley compared to the K-valley.

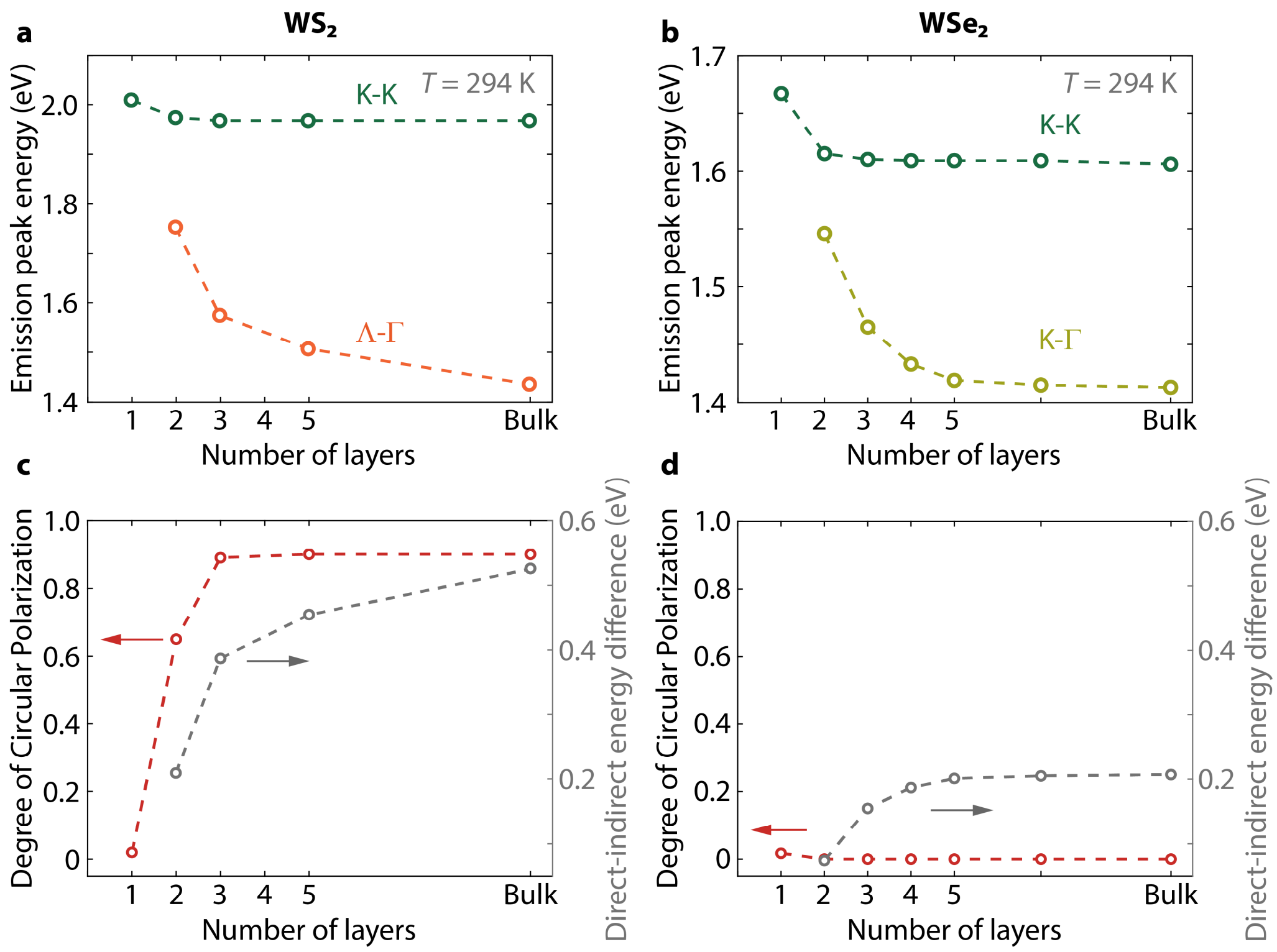

**Figure 2 | Different dependences of the spin-valley polarization of $WS_2$ and $WSe_2$ on the number of layers. a-b,** Photoluminescence peak positions of the transitions. **c-d,** Circular polarization of the K-K band maximum for different numbers of layers of $WS_2$ and $WSe_2$, respectively, under excitation at 2.040 eV. See text for similar results for $WSe_2$ excited close to resonance. In order to remove any possible spectral overlap with indirect transitions, we retrieve the polarization of the direct transition using Gaussian fitting of all the spectra detecting each circular polarization. The dashed lines are guides to the eye. All measurements at room temperature.

Next, we measure the change of polarization for a varying number of layers (Figures 2c and 2d, and Supplementary Figure 3). For $WS_2$, the polarization of the K-K transition quickly increases from mono- to trilayer, reaching a DOCP = 0.89 and saturating for thicker samples. For $WSe_2$, the polarization of the K-K transition remains absent for all thicknesses even when exciting closer to resonance (Supplementary Figure 4). As expected, the K-Γ and Λ-Γ transitions are unpolarized in all measurements (Supplementary Figures 2 and 3). We deconvolute the polarization contribution of each transition by fitting the spectra with Gaussian functions (see Methods). Thanks to this fit, we remove any contribution from the unpolarized indirect transition in our polarization analysis to retrieve the DOCP for the PL maximum of the direct transition alone.

The insensitivity of the polarization to thickness in $WSe_2$ is in clear contrast to the dependence in $WS_2$. As the main change in band structure with increasing thickness is a decrease in energy of the Λ-Γ transition ($E_{\Lambda\text{-}\Gamma}$), we can reasonably expect that an increasing difference between $E_{K\text{-}K}$ and $E_{\Lambda\text{-}\Gamma}$ could determine the increase in circular polarization in $WS_2$. To validate this hypothesis, however, we need to determine the conditions required for increasing the DOCP in $WSe_2$. Changing the temperature is a controllable way to perturb the band structure in both materials. Thus, we measure next the PL spectra and DOCP at lower temperatures and track the PL peak positions (Figure 3 and Supplementary Figure 4). In bilayer $WS_2$, the direct and indirect exciton peaks move to higher and lower energies with decreasing temperature, respectively (Figure 3a). In $WSe_2$, the situation is different. First, the K-Γ peak shifts to higher energy with decreasing temperature because the K-point is the conduction band minimum in this temperature range[31]. Below 160 K, the indirect peak starts moving to lower energies with decreasing temperature (Figure 3b), which is consistent with the indirect peak now arising from Λ-Γ transitions.

We describe the evolution of the peak energies with temperature (Figure 3a and b) using the Varshni equation[32]:

$$E_g(T) = E_g(0) - \frac{\alpha T^2}{\beta + T} \quad (1)$$

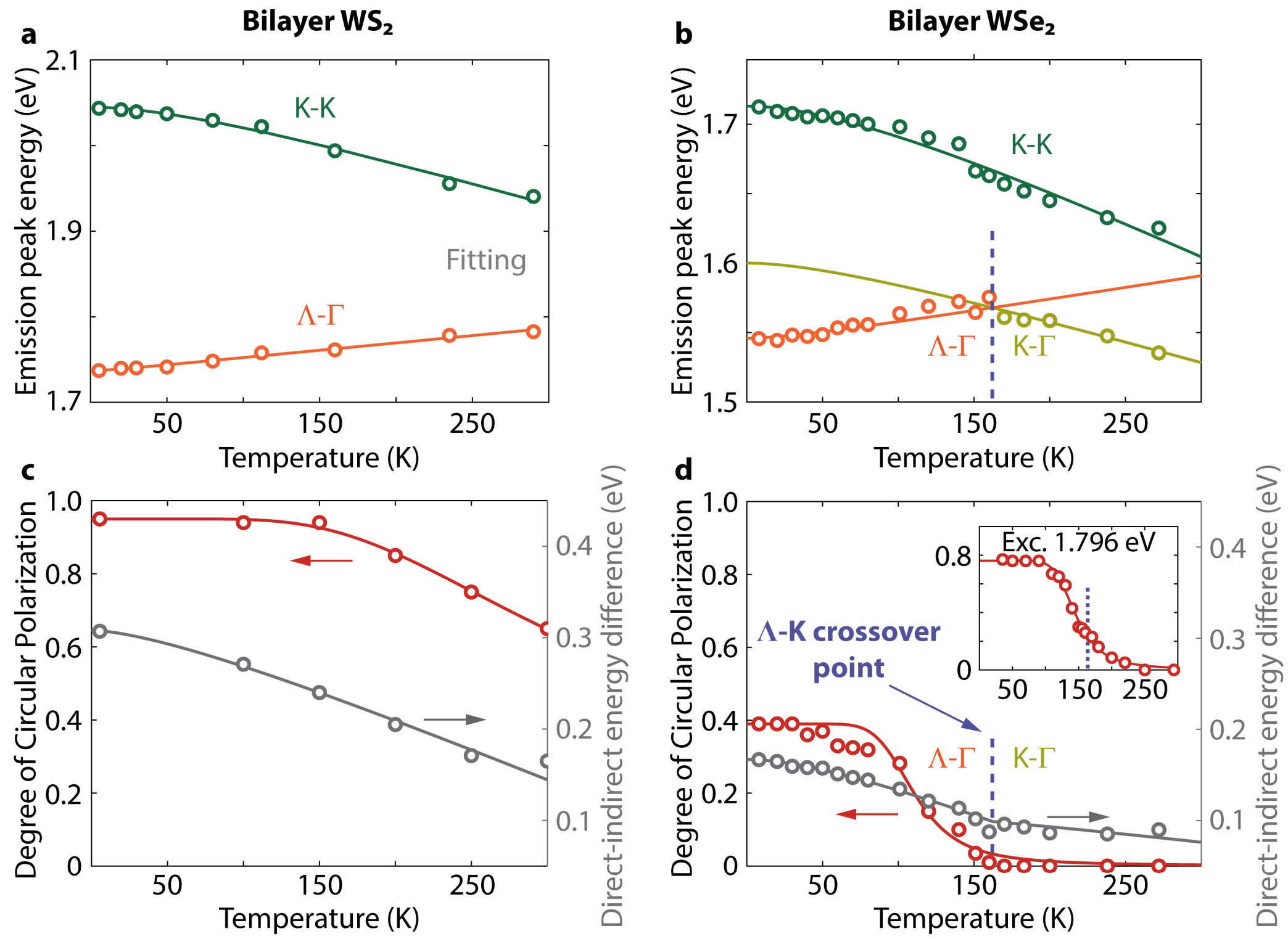

**Figure 3 | Relation between the indirect band gap character and spin-valley polarization. a-b,** Temperature dependence of the photoluminescence peak position for direct and indirect transitions for bilayer $WS_2$ and $WSe_2$. **c-d,** Temperature dependence of the circular polarization of the K-K transition maximum. The polarization of the K-K transition in $WSe_2$ starts increasing when the conduction band minimum shifts from K to Λ below the crossover temperature (blue dashed line). Inset: polarization under near-resonant excitation for $WSe_2$. Solid lines are fits as described in the text.

where $T$ is the temperature, $E_g(0)$ is the excitonic band gap at $T = 0$ K, and α and β are phenomenological fitting parameters. For the indirect exciton in $WSe_2$, we use two separate Varshni equations for the high- and low-temperature regimes due to the change in indirect transition character at $T = 160$ K. We list the fitting parameters in Table 2. The parameter $\alpha$ describes the band gap change with temperature due to thermal expansion of the lattice. The values of $\alpha$ for the K-K transition and the Λ-Γ transition are equal for both materials, demonstrating their similar dependence of band structure on temperature. However, bilayer $WSe_2$ has a smaller band gap than $WS_2$. As a result, the K-K and Λ-Γ transitions will be closer in energy in $WSe_2$. Therefore, the crossover of the indirect transitions occurs at a lower temperature in $WSe_2$ than in $WS_2$.

**Table 2** | Fitting parameters using the Varshni equation for the temperature dependence of the different transitions and materials in Figures 3a and 3b.

| **Material / Transition** | | $E_g(0)$ **(eV)** | $\alpha$ **(meV/K)** | $\beta$ **(K)** |
|---|---|---|---|---|
| **$WS_2$** | K-K | 2.045 | 0.530 | 118.9 |
| | Λ-Γ | 1.737 | -0.172 | 12.5 |
| **$WSe_2$** | K-K | 1.713 | 0.530 | 139.3 |
| | K-Γ | 1.600 | 0.316 | 96.6 |
| | Λ-Γ | 1.546 | -0.172 | 44.1 |

The measured polarization rises with increasing number of layers in $WS_2$ because it corresponds to a higher K-Λ energy separation of the conduction bands. Similarly, this energy separation also increases with decreasing temperature. Consequently, for bilayer $WS_2$ the circular polarization increases with decreasing temperature as well (Figure 3c). At temperatures from 300 to 160 K, valley polarization remains absent in $WSe_2$ when excited off resonance at 2.040 eV. In this temperature range, the $E_{K\text{-}K}$-$E_{K\text{-}\Gamma}$ separation does not vary substantially because both peaks shift to higher energy with decreasing temperature (gray points in Figure 3d). Below $T$ = 160 K, the indirect transition changes from K-Γ to Λ-Γ. Simultaneously, the polarization of the K-K transition starts to increase and saturates at low temperatures (Figure 3d, including higher polarization under near resonant excitation). At these low temperatures, bilayer $WSe_2$ behaves like bilayer $WS_2$ because their indirect transitions have now both Λ-Γ character, as evidenced by their similar Varshni dependences.

Our polarization values at low temperature ($T$ ~ 10 K) are consistent with previous measurements using similar excitation energies for both materials.[20,33] Reaching a DOCP of 0.39 at 35 K in bilayer $WSe_2$, despite exciting 320 meV away from the K-K transition, demonstrates the critical role of the Λ-valley in establishing the robust spin-valley polarization in few-layer $WS_2$ and $WSe_2$.

To compare the temperature dependence of polarization in few-layer $WS_2$ and $WSe_2$, we fit the DOCP as a function of temperature (Figures 3c and 3d) using the expression

$$\mathrm{DOCP} = \frac{P_0}{1+2\tau_{PL}/\tau_v}, \qquad (2)$$

which takes into account the K-K exciton transition rate ($1/\tau_{PL}$) and the K-K' intervalley scattering rate ($1/\tau_v$).[1] $P_0$ is the initial polarization before scattering takes place, for which we use the maximum DOCP at the lowest measured temperature. We assume that excitons follow a Boltzmann distribution for the ratio $\tau_{PL}/\tau_v = c\, exp(-\Delta E/k_B T)$,[20] where $c$ is a constant and $\Delta E$ is an activation energy needed to undergo K-K' intervalley scattering (see Supplementary Table 1 for fitting parameters).

Next, we explicitly demonstrate that the polarization of the direct K-K transition in $WSe_2$ depends on the K-Λ energy difference in the conduction band, $\Delta E_{\text{K-Λ}}$. First, we retrieve $\Delta E_{\text{K-Λ}}$ as a function of temperature from fitting Figure 3b as $\Delta E_{\mathrm{K}-\Lambda} = E_{\mathrm{K}-\Gamma} - E_{\Lambda-\Gamma}$ (Figure 4a). The polarization of the K-K transition starts increasing when $\Delta E_{\text{K-Λ}}$ becomes positive (Figure 4b). Previously, the Γ-hill has been suggested to be involved in K-K' intervalley scattering by slowing down the scattering of holes from K to K' in bilayer $WS_2$ compared to monolayer $WS_2$.[34] That hypothesis is not consistent, however, with the absence of valley polarization in $WSe_2$ at higher temperatures, where the dominant indirect transition is K-Γ. Instead, the Λ-valley could play a similar role in the scattering of electrons. A comparison of Δ$E_{\text{K-}\Lambda}$ in the conduction band and Δ$E_{\Gamma\text{-K}}$ in the valence band ($\Delta E_{\Gamma-K} = E_{K-K} - E_{K-\Gamma}$) further supports the relevance of the Λ-valley for polarization (Figure 4a). The energy difference in the valence band Δ$E_{\Gamma\text{-K}}$ is already far above the thermal energy at room temperature, so its weak increase at lower temperatures cannot influence polarization substantially. On the other hand, Δ$E_{\text{K-}\Lambda}$ is similar to the thermal energy near the Λ-K crossover. Then, excitons will populate both Λ-Γ and K-Γ states, resulting in the weaker polarization increase before the Λ-K crossover (Figure 3d). Our results thus highlight the important role of the Λ-valley in protecting spin-valley polarization in few-layer semiconductors.

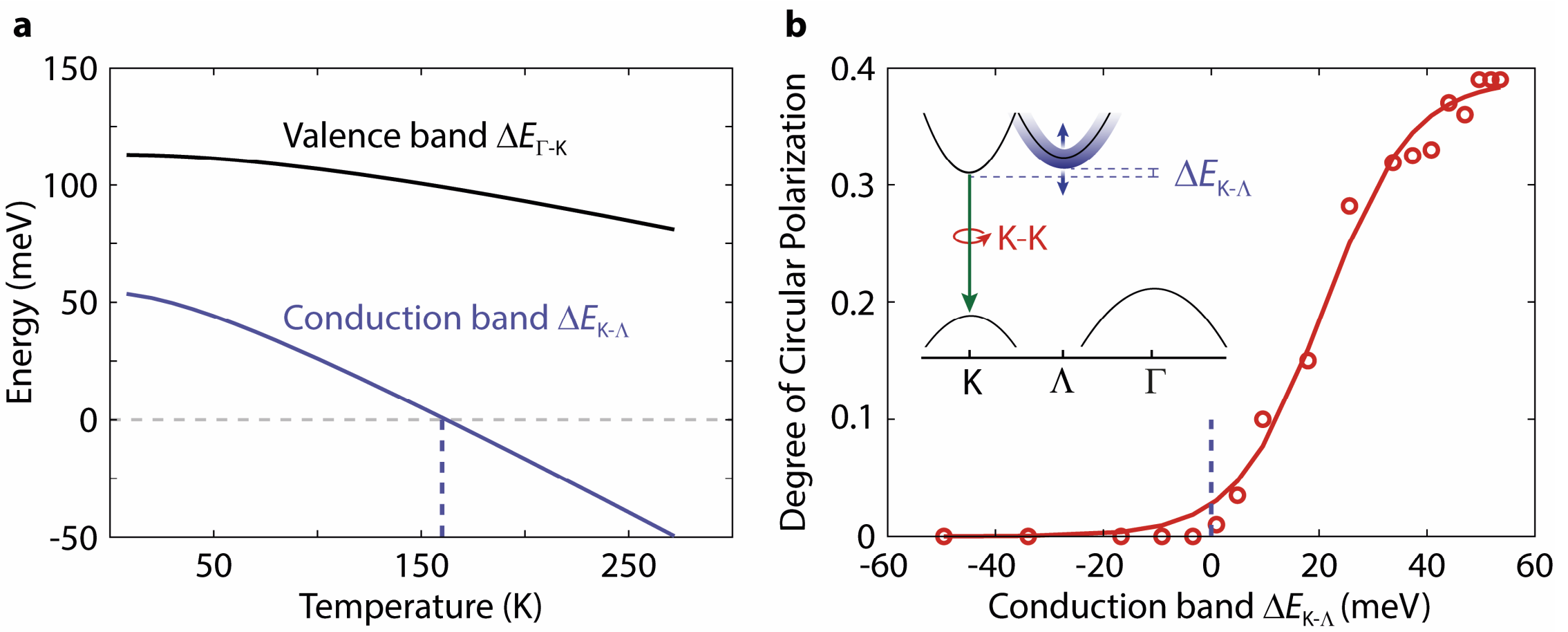

**Figure 4 | The polarization of the direct K-K transition in $WSe_2$ depends on the energy difference between the conduction band points of the indirect transitions. a,** Temperature dependence of Δ$E_{\text{K-}\Lambda}$ in the conduction band and Δ$E_{\Gamma\text{-K}}$ in the valence band obtained from the three fits in Figure 3b. **b,** Polarization of the K-K transition excited with 2.040 eV as a function of the Δ$E_{\text{K-}\Lambda}$ conduction band difference. Inset: schematic of the band diagram responsible for the changes in K-K polarization determined by the value of Δ$E_{\text{K-}\Lambda}$.

### Mechanisms for spin-valley polarization in few-layer semiconductors

To better illustrate the appearance of polarization below the crossover temperature between indirect transitions, we depict how the band structure changes for three different temperatures (Figure 5). When $T > 160$ K, the indirect transition is K-Γ and polarization is absent for the K-K transition (Figure 5, left). As the temperature decreases, Λ moves to an energy similar to K. There is an intermediate temperature range where both K-Γ and Λ-Γ transitions contribute to the indirect emission and the K-K polarization starts to increase (Figure 5, center). The overlap in emission from both indirect transitions is evident from the fitting of the indirect spectral band, where two Gaussians are necessary. Finally, as $\Delta E_{K\text{-}\Lambda}$ increases, only the Λ-valley contributes to indirect emission resulting in a faster increase of polarization with decreasing temperature (Figure 5, right).

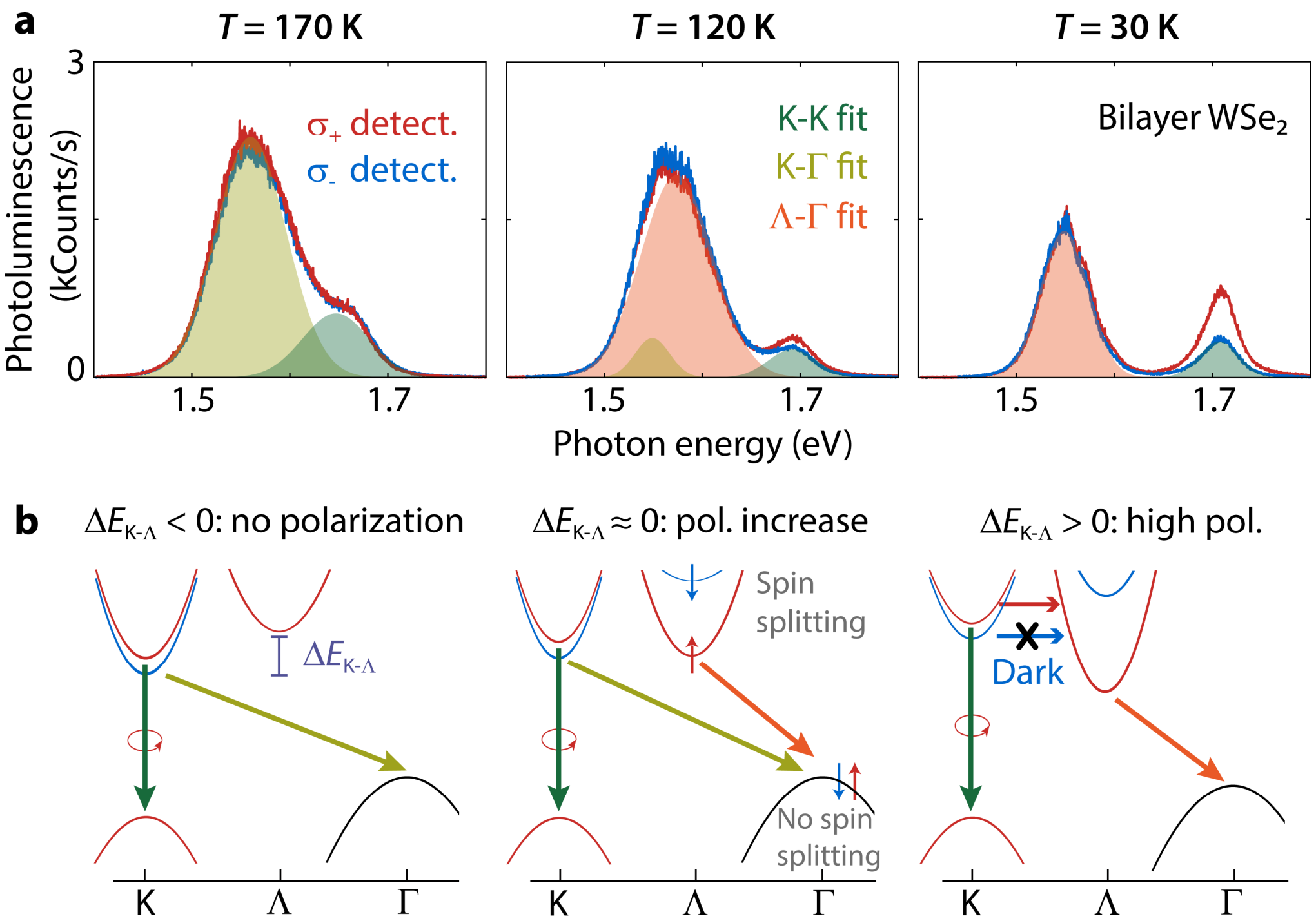

**Figure 5 | Effect of the K-Λ crossover on the K-K transition polarization for bilayer $WSe_2$. a,** Photoluminescence spectra and **b,** band structure schematics illustrating the conditions leading to the appearance of polarization at three different temperatures. The emission spectra include fits to a Gaussian function for the detected $\sigma_+$-polarization spectra. At $T = 170$ K (left), there is no polarization and the Λ-valley is not yet involved in the transitions. At this temperature, the K-Γ transition can deplete the spin-down K-valley because there is no spin splitting in the Γ-hill. At $T = 120$ K (center), a small spin-valley polarization appears as the Λ-valley now takes part in the

optical transitions. Here, two Gaussians are necessary for fitting the indirect PL peak. At $T = 30$ K (right), only the Λ-valley contributes to the indirect emission resulting in higher polarization. Due to spin splitting in the Λ-valley, the spin-down K-valley is no longer depleted through indirect optical transitions and becomes dark. All spectra were acquired using 2.040 eV excitation.

We discuss next how the Λ-valley could lead to a reduced intervalley scattering rate. We focus on two mechanisms for intervalley scattering and how they compare to our results:

(I) Intervalley scattering by phonons.[35–37]

(II) Intervalley scattering by the long-range exchange mechanism.[16,38,39]

The coupling between excitons and phonons can cause the spin to flip, but this process also requires phonons with the right momentum. A strong exciton-phonon coupling promotes spin-flipping.[40] We extract the exciton-phonon coupling using the the O'Donnell equation[41] to fit the PL peaks in Figures 3a and 3b. The results indicate that Λ-Γ excitons have a lower exciton-phonon coupling strength (Supplementary Section 2.2) and are thus the least likely to undergo intervalley scattering by phonons. Additionally, intervalley scattering by phonons can occur much faster for electrons due to the low spin splitting of the conduction band at the K-point enabling faster spin-flipping.[35] Due to its lower energy and the very large spin splitting[42], the Λ-valley could introduce a very efficient 'trap' preventing electrons scattered by phonons from reaching the K'-valley. Thus, K-K' intervalley scattering by phonons could be slowed down significantly when the Λ-valley is the conduction band minimum. Intervalley scattering by phonons is thus consistent with the polarization trends illustrated in Figure 5.

Alternatively, the long-range exchange mechanism can cause a K-K exciton to undergo intervalley scattering by recombining and exciting an exciton at the K'-valley. Compared to phonons, this mechanism does not require any additional momentum. Since the process occurs more efficiently for excitons of higher kinetic energy, the Λ-valley slows down the K-K' intervalley scattering rate by trapping K-K excitons of higher kinetic energy more efficiently. From the fitting of DOCP as a function of temperature with Equation 2, we note that the values of $\Delta E$ (Supporting Table 3) do not match with the phonon energy required for K-K' intervalley scattering.[43,44] $\Delta E$ could thus correspond better with the excess energy required by the long-range exchange mechanism. Nevertheless, such a simple fit likely does not take into account all of the temperature-dependent material parameters affecting spin-valley polarization.[17]

Additionally, spin-forbidden dark excitons with lower energy than the bright excitons[45] can increase polarization. Dark excitons have previously been attributed to the robust spin-valley polarization in monolayer tungsten systems with respect to both temperature and excitation energy.[47] Our results show the presence of dark excitons in Supplementary Section 2.3 and we estimate a bright-dark splitting $E_{BD}$ = 37.9 meV, which is in agreement with the value in monolayer $WSe_2$.[50] As there is no K-K' intervalley exchange interaction for dark excitons[16], they can act as a reservoir for the bright exciton valley polarization[47]. On the other hand, the K-Γ transition depopulates the dark exciton reservoir contributing to depolarization because there is no spin splitting at Γ.[21] This situation is consistent with the low polarization at temperatures above the Λ-K crossover (Figure 5, left). In the temperature range where $E_\Lambda < E_K$, the K-K dark exciton reservoir is restored due to the spin splitting at Λ[21,42], resulting in a high and robust polarization even for off-resonant excitation (Figure 5, right).

Finally, we show for completeness that the dependence of polarization on the number of layers is also in agreement with the mechanism in Figure 5, where polarization is controlled by $\Delta E_{K\text{-}\Lambda}$. For $WS_2$, there is a significant difference in polarization between bilayers and trilayers at room temperature (DOCP ≈ 0.65 and 0.95 in Figure 2c). The addition of one layer shifts the Λ-valley to lower energy, while the K-valley remains nearly constant (Figure 2a), thus increasing $\Delta E_{K\text{-}\Lambda}$. To confirm this behavior in $WSe_2$ as well, we perform temperature- and polarization measurements on $WSe_2$ from one to four layers showing an increase of polarization with number of layers (Supplementary Figure 7). The results are consistent with the expectation of a smaller $\Delta E_{K\text{-}\Lambda}$ energy separation at elevated temperatures with increasing thickness, resulting in the Λ-K crossover occurring at a higher temperature. The polarization at $T = 35$ K is much higher in bilayer $WSe_2$ excited off resonance than in monolayer $WSe_2$ excited near resonance (DOCP ≈ 0.79 compared to 0.23) despite the additional detuning. This polarization difference between bilayer and monolayer $WSe_2$ is consistent with the Λ-valley offering additional protection for K-K' intervalley scattering by phonons in bilayers.

## Conclusion

In summary, we have demonstrated the impact of the Λ-valley on spin-valley polarization in $WS_2$ and $WSe_2$ through temperature- and polarization-resolved photoluminescence measurements. By varying the temperature and the number of layers, the position of the conduction band Λ-valley changes relative to the K-valley. We show that the conduction band Λ-K energy difference controls the spin-valley polarization of the direct K-K transition resulting in robust polarization. In bilayer $WSe_2$, we correlate the appearance of polarization with the crossover between indirect transitions below $T = 160$ K, when the Λ-valley becomes the conduction band minimum. The polarization increases with the energy difference between the K- and Λ-valleys. This observation underlines the importance of the Λ-valley in blocking K-K' intervalley scattering to stabilize polarization.

Our results introduce the critical role of indirect optical transitions in spin-valley polarization in few-layer semiconductors, contributing in particular to the understanding of the exceptionally high spin-valley polarization in few-layer $WS_2$. For $WS_2$, the energy of the Λ-valley is already lower than the K-valley at room temperature. Polarization increases with the number of $WS_2$ layers because of the higher K-Λ energy difference. The Λ-valley thus determines the contrast between the high polarization in few-layer $WS_2$ and low polarization for monolayer $WS_2$. It also causes the contrast between few-layer $WS_2$ and $WSe_2$ at room temperature. The protection of polarization by the emergence of an indirect transition is a striking manifestation of interlayer interactions at the sub-nanometer scale. The control of the band structure and its indirect transitions by changing the interlayer distance (*e.g.*, using strain or pressure), tuning the band gap (*e.g.*, via electrical gating), or through hetero- or homostructures opens a route to manipulate the entanglement of the spin, valley, and layer indexes for actively tunable *valleytronics*.

**Methods**

**Sample fabrication.** We deposit $WS_2$ and $WSe_2$ microcyrstals onto $SiO_2$/Si (285-nm thick $SiO_2$) substrates by mechanical exfoliation from synthetic, bulk 2H crystals (HQ Graphene). We first determine the thickness of the flakes by optical contrast microscopy[51] and by considering the energy of the indirect exciton emission in photoluminescence spectra[52]. After optical measurements, we confirmed the thickness by atomic force microscopy.

**Optical measurements.** We carry out photoluminescence measurements using a microscope in epi-fluorescence geometry (objective lens: Nikon CFI Plan Fluor ELWD 40x, NA = 0.6). We typically excite the microcrystals with a continuous-wave laser with photon energy 2.040 eV and a power of 12.2 μW before the objective lens, ensuring a power density in the linear response range for the TMDs. For $WSe_2$, we also used a continuous-wave 1.796 eV laser with a power of 75.6 μW. For one measurement with bilayer $WSe_2$, we used a supercontinuum laser (Fianium SC400, pulse duration ~50 ps) with an acousto-optical filter tuned to 1.681 eV and 1.7 μW. To control the circular polarization in excitation, we employ a Babinet-Soleil compensator (Melles Griot) and a Stokes polarimeter (PolSNAP, Hinds Instruments) at the sample position to ensure circular polarization of the incident laser beam in the absence of the objective. In the detection path, we use a non-polarizing beamsplitter (21014 Silver Non-Polarizing 50/50 bs, Chroma), and then either two 615-nm longpass filters (ET615LP Chroma), two 700-nm longpass filters (FELH0700, Thorlabs), or one 750-nm longpass filter (FELH0750, Thorlabs). For emission polarization analysis, we combine a quarter-wave plate (achromatic quarter-wave retarder, 600-1200 nm, Bernhard Halle) and a wire-grid polarizer (WP25M-UB, Thorlabs). After coupling into an optical fiber with core size 50 μm, we record PL spectra with an Andor Shamrock 303i spectrometer and an Andor Newton 970 EMCCD camera. For low-temperature measurements, we use a liquid helium flow cryostat (Oxford Instruments MicrostatHiRes) pumped to ultra-high vacuum.

**Acknowledgments**. We thank Marcos H. D. Guimarães for useful discussions. This work was financially supported by the Netherlands Organisation for Scientific Research (NWO) through Gravitation grant "Research Centre for Integrated Nanophotonics" (024.002.033) and an NWO START-UP grant (740.018.009) and the Innovational Research Incentives Scheme (VICI Grant nr. 680-47-628). Shaojun Wang was supported by the Starting Grant of Soochow University (Q415900120) and the Priority Academic Program Development (PAPD) of Jiangsu Higher Education Institutions.

# Supplementary Information

# Impact of indirect transitions on valley polarization in $WS_2$ and $WSe_2$

Rasmus H. Godiksen[1], Shaojun Wang[1,2], T.V. Raziman[1],

Jaime Gómez Rivas[1], Alberto G. Curto[1,3,4*]

*[1]Dep. Applied Physics and Institute for Photonic Integration,*

*Eindhoven University of Technology, Eindhoven, The Netherlands*

*[2]MOE Key Lab. of Modern Optical Technologies and Jiangsu Key Lab. of Advanced Optical Manufacturing Technologies, School of Optoelectronic Science and Engineering, Soochow University, Suzhou 215006, China*

*[3]Photonics Research Group, Ghent University-imec, Ghent, Belgium*

*[4]Center for Nano- and Biophotonics, Ghent University, Ghent, Belgium*

* Corresponding author: A.G.Curto@TUe.nl

## Supplementary Section 1: Room temperature photoluminescence

### 1. Resonant excitation of bilayer $WSe_2$

Throughout our study, we report polarization values measured with excitation at a constant photon energy of 2.040 eV, which is close to resonance for bilayer $WSe_2$ . Here, we check that bilayer $WSe_2$ does not show spin-valley polarization either under resonant excitation conditions (Supplementary Figure 1). We used an excitation energy of 1.681 eV, which is close to the $WSe_2$ bilayer K-K exciton emission around 1.62 eV.

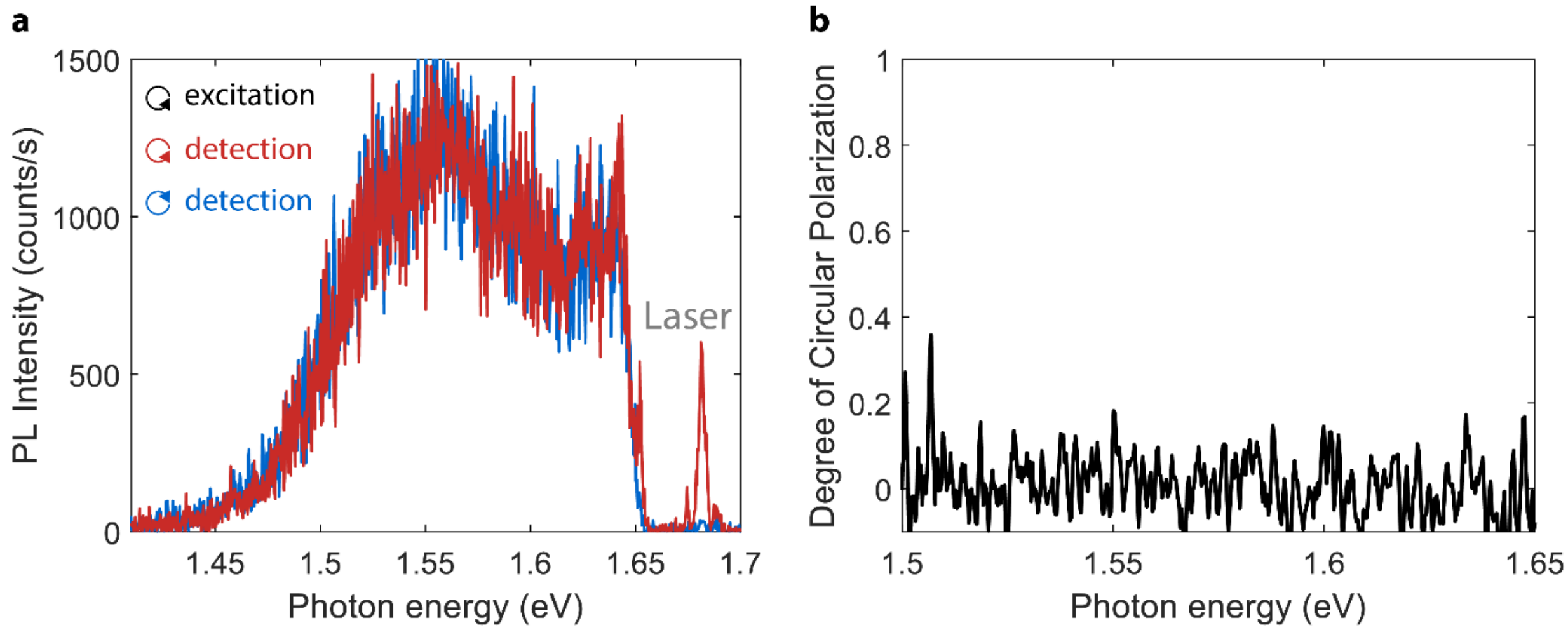

**Supplementary Figure 1. a,** Polarization-resolved PL spectrum for bilayer $WSe_2$ under near-resonant excitation (1.681 eV) at room temperature. **b,** Degree of circular polarization as a function of photon energy. No DOCP is measurable.

### 2. Thickness-dependent polarization

We determine the thickness of our $WS_2$ and $WSe_2$ samples using a combination of reflection contrast microscopy, atomic force microscopy, and photoluminescence measurements. Similarly, for both $WS_2$ and $WSe_2$, the monolayers show bright emission due to their direct band gap. At room temperature, their emission spectrum shows a single peak. When increasing the thickness, a second peak emerges, which shifts to lower energy with an increasing layer thickness (Supplementary Figure 2). Only $WS_2$ shows an increase in the DOCP with thickness, whereas in $WSe_2$ the emission remains unpolarized for all thicknesses when excited with 2.04 eV (Supplementary Figure 3) and 1.796 eV (Supplementary Figure 4).

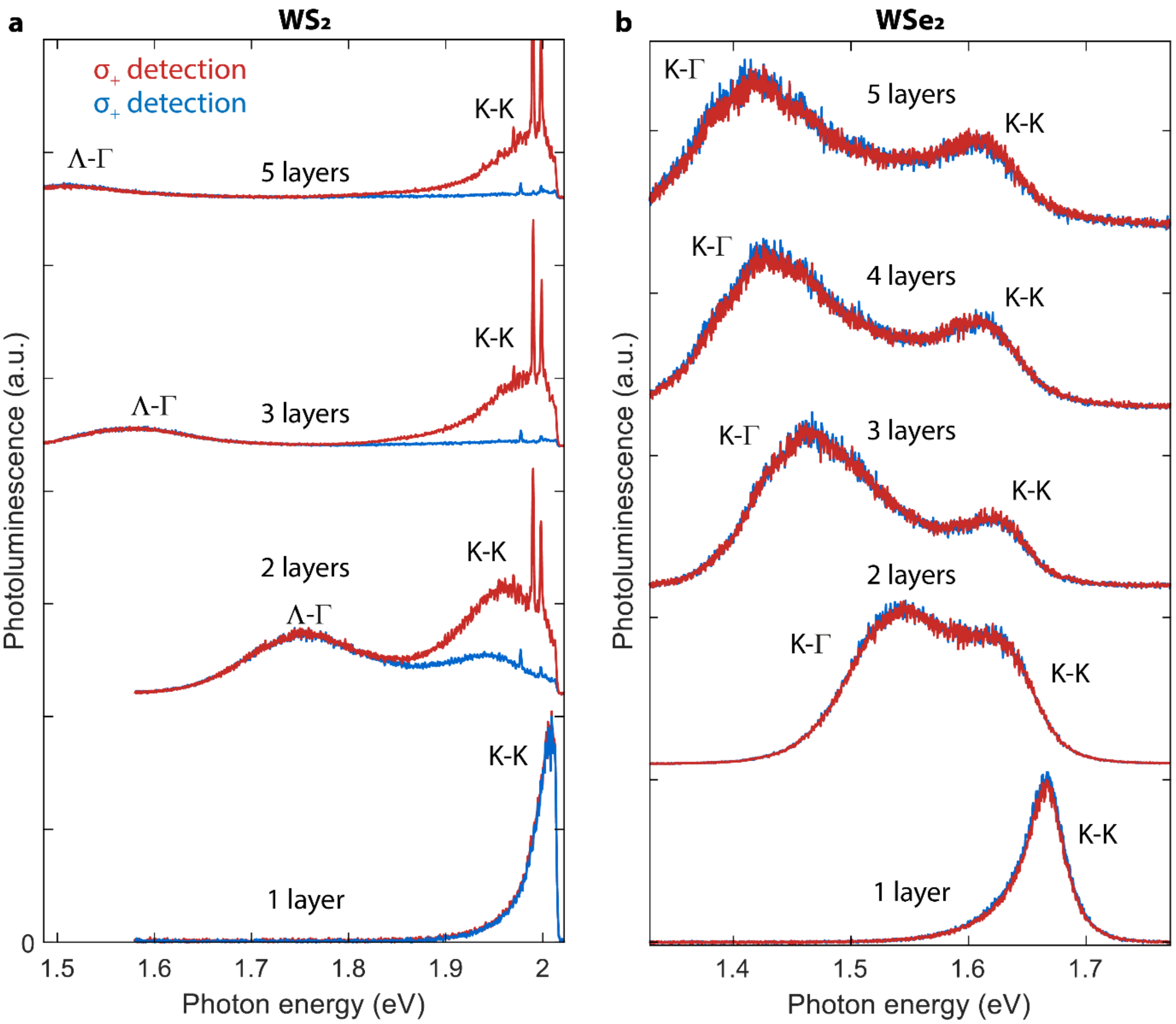

**Supplementary Figure 2.** Polarization-resolved PL spectra at room temperature for different thicknesses. **a,** $WS_2$. **b,** $WSe_2$. Spectra are vertically shifted by a constant for clarity.

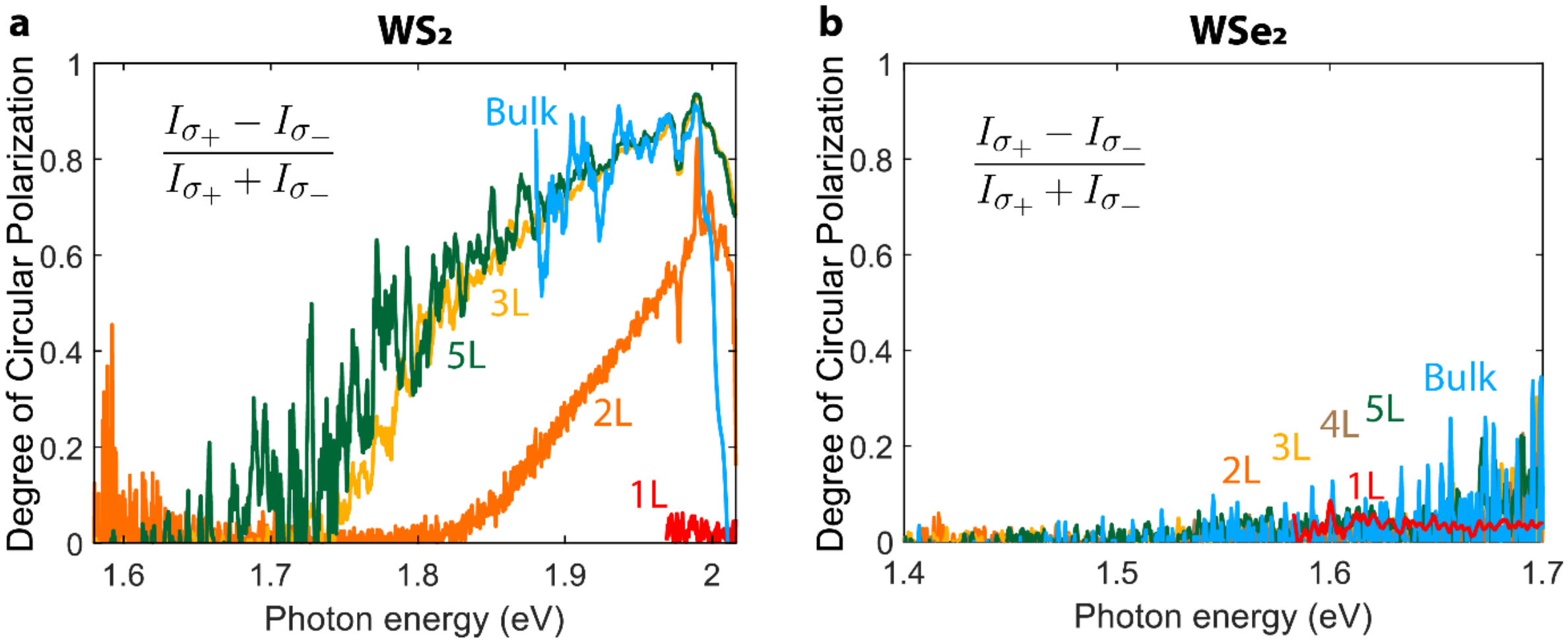

**Supplementary Figure 3.** Degree of circular polarization at room temperature as a function of emission wavelength for a set of different thicknesses for **a,** $WS_2$, and **b,** $WSe_2$.

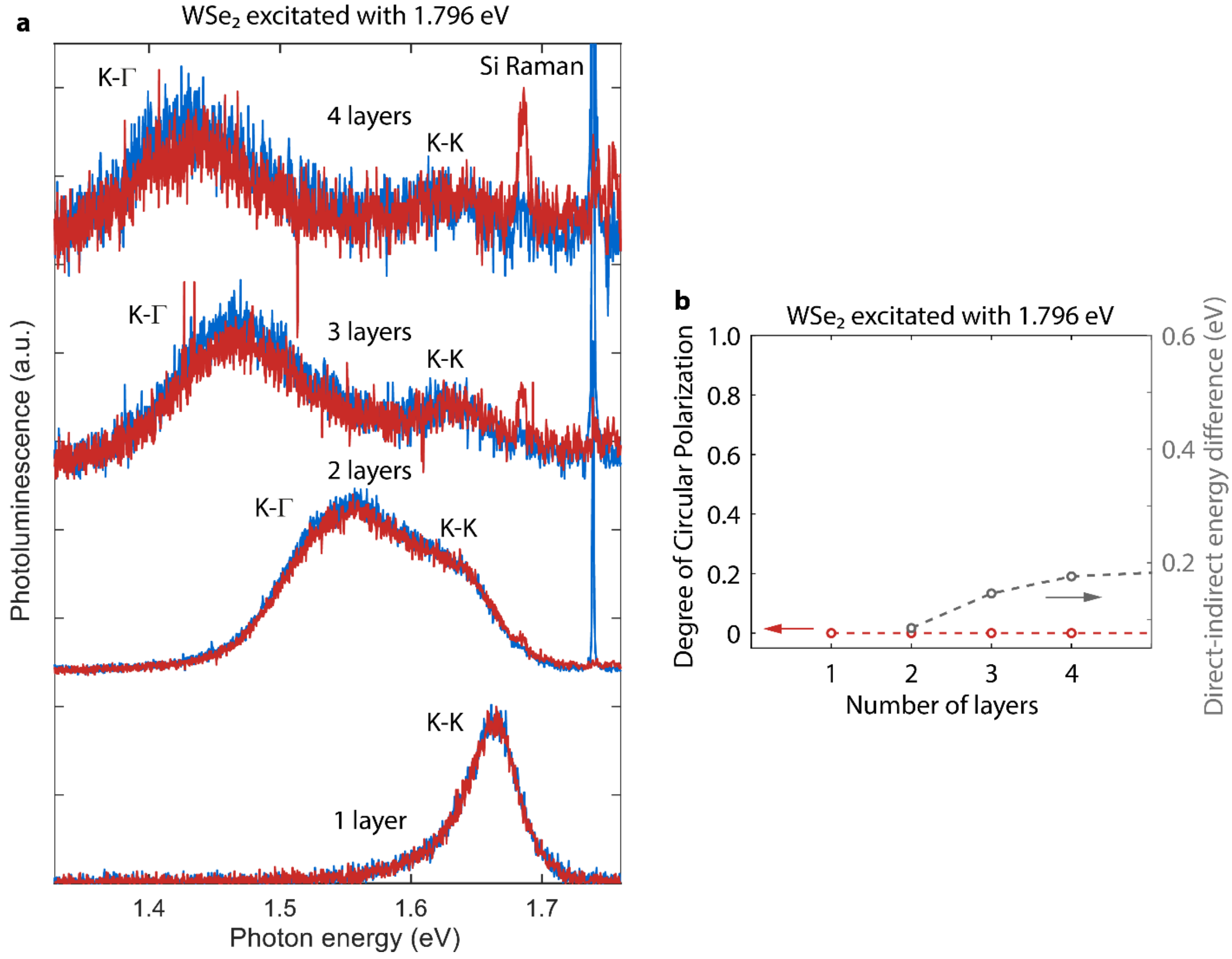

**Supplementary Figure 4.** Polarization-resolved PL spectra at room temperature for different thicknesses of $WSe_2$ excited with 1.796 eV. **a,** Thickness-dependent spectra. **b,** Thickness vs. DOCP and the direct-indirect energy difference for the spectra in **a**.

## Supplementary Section 2: Temperature-dependent photoluminescence

### 1. Bilayer photoluminescence spectra

When the temperature decreases, the two photoluminescence peaks shift with temperature (Supplementary Figure 4). In bilayer $WS_2$, the polarization also increases. However, in bilayer $WSe_2$, polarization only appears below $T = 160$ K (Supplementary Figure 5).

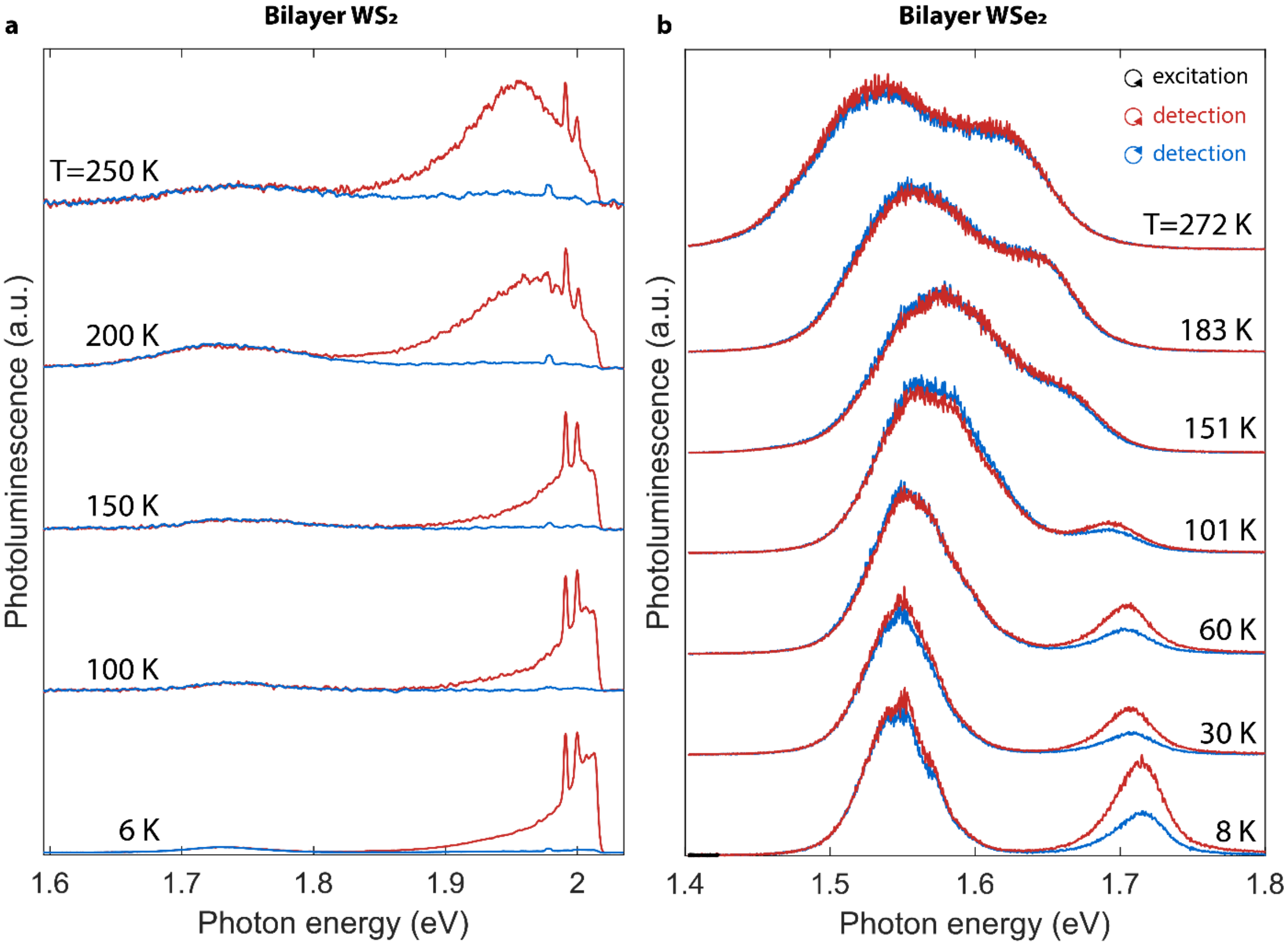

**Supplementary Figure 5.** Polarization-resolved PL spectra at different temperatures for bilayer samples. **a,** $WS_2$. **b,** $WSe_2$. Spectra are vertically shifted by a constant for clarity.

**Supplementary Table 1.** Fitting parameters obtained using Equation 2 in the main text in Figure 3. We contained the factor of 2 in the denominator of Equation 2 in the fitting parameter, $c$.

| Material | $c$ | $\Delta E$ (meV) |
|---|---|---|
| $WS_2$ | 0.12 | 74.6 |
| $WSe_2$ | 0.076 | 72.5 |

## 2. Fitting using the O'Donnel equation

We fit the peak position as a function of temperature using two equations. Fitting using the Varshni equation1 was presented in the main text. Here, we fit the peak position using the O'Donnell equation[1]

$$E_g(T) = E_g(0) - S\langle\hbar\omega\rangle[\coth(\langle\hbar\omega\rangle/2k_BT) - 1] \quad \text{(S1)}$$

where $T$ is the temperature, $E_g(0)$ is the excitonic band gap, $S$ is the Huang-Rhys factor, $\langle\hbar\omega\rangle$, is an average phonon energy, and $k_B$ is the Boltzmann constant. The obtained fitting parameters are listed in Supplementary Table 2. Fitting using the O'Donnell equation yields as good a fit as with the Varshni equation, i.e., $R^2 = 0.9999$ when comparing the two fits. The main variation one might encounter between these two fitting methods will be expressed mainly in the range T=0-20 K, where we do not have several data points, as the band gap energy varies less in this range. The O'Donnell equation has a more profound theoretical background, and its fitting parameters are more well defined[1]. The Huang-Rhys factor, $S$, describes the exciton-phonon coupling strength of a certain transition. Comparing the values for each transition in Supplementary Table 2, we note that the exciton-phonon coupling strength is much larger for transitions that involve electrons in the K-valley compared to the Λ-valley. Similarly, the average phonon energy is also smaller for Λ-Γ excitons, suggesting that Λ-Γ excitons are more resistant to scattering by phonons.

**Supplementary Table 2.** Fitting parameters obtained using Equation S1 with the experimental data in Figure 3a-b.

| Material / Transition | | $E_g(0)$ (eV) | $S$ (-) | $\langle \hbar\omega \rangle$ (meV) |
|---|---|---|---|---|
| $WS_2$ | K-K | 2.045 | 2.979 | 14.6 |
| | Λ-Γ | 1.737 | 0.997 | 2.0 |
| $WSe_2$ | K-K | 1.713 | 2.957 | 16.5 |
| | K-Γ | 1.600 | 1.791 | 12.4 |
| | Λ-Γ | 1.546 | 0.991 | 6.3 |

### 3. Dark ground state in bilayer $WSe_2$

In W-based monolayers, the dark excitons lie lower in energy than the bright excitons and transitions between the lowest conduction band and the top valence band at K is spin-forbidden (dark K-K exciton) due to spin splitting[2]. As evidence for bright-dark excitons in bilayer $WSe_2$, we observe a decrease of the K-K intensity with decreasing temperature consistent with reduced thermalization from dark to bright excitons[2–4] (Supplementary Figure 6). We fit the measured integrated PL intensity as a function of temperature to the expression $I_{PL}(T)/I_{PL}(0) - 1 = C\ exp(-E_D/k_B T)$, where $I_{PL}(T)$ is the measured intensity as a function of temperature, $I_{PL}(0)$ is the intensity at $T = 0$ K, $C$ is a constant, $k_B$ is the Boltzmann constant, and $E_D$ is the characteristic energy barrier that defines the slope of the emission. From the fit, we obtain $E_D = 37.9$ meV, which is in good agreement with the bright-dark exciton splitting in monolayer $WSe_2$[5]. We expect a similar value for bilayer $WSe_2$ due to the limited effect of layer-layer interactions on the band structure near the K-point of the Brillouin zone.

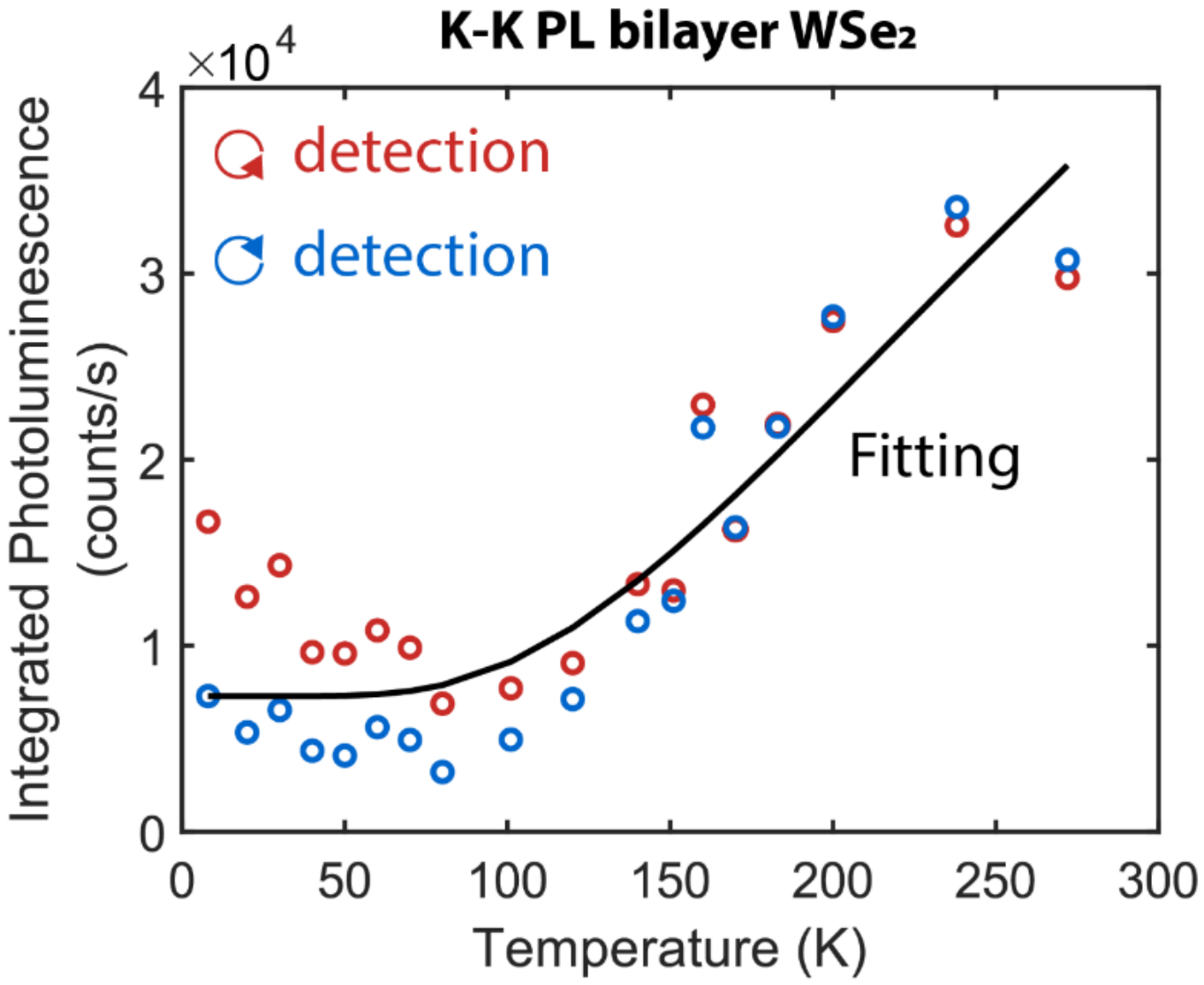

**Supplementary Figure 6.** Spectrally integrated PL of the A exciton emission as a function of temperature for both circular polarizations when excited with a 2.04 eV laser. The drop in emission intensity with temperature is consistent with a dark exciton ground state. The fitting is described in the text.

## 4. Temperature-dependent polarization with varying thickness in $WSe_2$

For a fixed temperature, if we increase the $WSe_2$ thickness to three or four layers, the K-Λ conduction band difference should become smaller. Similarly, the onset of an increase in DOCP should occur at a higher temperature compared to a bilayer. We confirm this trend by measuring the emission DOCP for three and four layers of $WSe_2$ and comparing it as a function of temperature to that of a bilayer (Supplementary Figure 7).

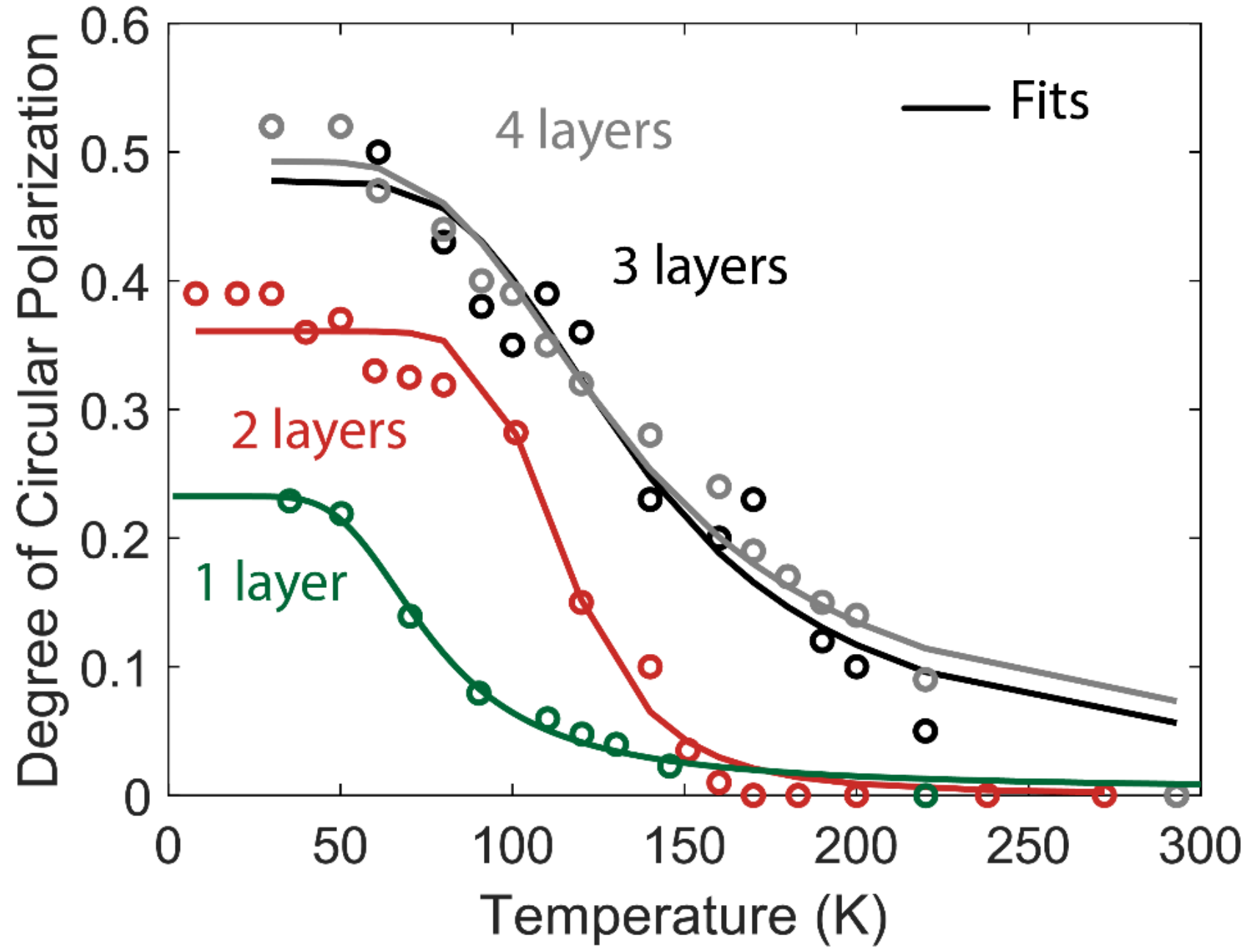

**Supplementary Figure 7.** Temperature-dependent DOCP measurements for 1, 2, 3, and 4 layers of $WSe_2$ showing an increase in the onset temperature of DOCP with increasing layer thickness. The 2, 3, and 4 layer data was acquired using 2.04 eV excitation. The 1 layer data was acquired using 1.796 eV excitation. The fits are made by assuming a Boltzmann distribution for the K-K' intervalley scattering, see details in the main text.